\documentclass[12pt]{article}
\begin{document}
\newcommand{\be}{\begin{equation}}
\newcommand{\ee}{\end{equation}}
\newcommand{\al}{\alpha}
\newcommand{\bt}{\beta}
\newcommand{\lm}{\lambda}
\newcommand{\bea}{\begin{eqnarray}}
\newcommand{\eea}{\end{eqnarray}}
\newcommand{\gm}{\gamma}
\newcommand{\Gm}{\Gamma}
\newcommand{\dl}{\delta}
\newcommand{\Dl}{\Delta}
\newcommand{\ep}{\epsilon}
\newcommand{\kp}{\kappa}
\newcommand{\Lm}{\Lambda}
\newcommand{\om}{\omega}
\newcommand{\pa}{\partial}
\newcommand{\dd}{\mbox{d}}
\newcommand{\MS}{\mbox{MS}}
\newcommand{\nn}{\nonumber}
\newcommand{\uk}{\underline{k}}

\begin{flushright}
  DESY/01-086\\
  TTP01-15\\
  June 2001\\
\end{flushright}

\vspace{1.0cm} 
\begin{center}{\LARGE
Next-to-Next-to-Leading Logarithms in Four-Fermion \\[1ex]
Electroweak Processes at High Energy}
\end{center}

\vspace{0.5cm}

\begin{center}{
  {\large J.H.~K\"uhn} $^a$,\,
  {\large S.~Moch} $^a$,\,
  {\large A.A.~Penin} $^{b,c}$,\, and  
  {\large V.A.~Smirnov} $^d$\\
\vspace{1.0cm} 
  $^a${\small {\em Institut f{\"u}r Theoretische Teilchenphysik,
  Universit{\"a}t Karlsruhe}}\\
  {\small {\em 76128 Karlsruhe, Germany}}\\[2mm]
  $^b${\small {\em II. Institut f{\"u}r Theoretische Physik,
  Universit{\"a}t Hamburg}}\\
  {\small {\em  22761 Hamburg, Germany}}\\[2mm]
  $^c${\small {\em Institute for Nuclear
  Research of Russian Academy   of Sciences,}}\\
  {\small {\em 117312 Moscow, Russia}}\\[2mm]
  $^d${\small {\em  Nuclear Physics
  Institute of Moscow State University}}\\
  {\small {\em  119899 Moscow, Russia}}
}
\end{center}

\vspace{0.5cm} 

\begin{abstract}
We sum up 
the next-to-next-to-leading logarithmic virtual
electroweak corrections to the high energy  asymptotics of the
neutral current four-fermion processes for light fermions
to all orders in the coupling constants
using the evolution equation approach. From this 
all order result we derive finite order expressions 
through next-to-next-to leading order for the total cross 
section and various asymmetries. We observe an amazing
cancellation between the sizable leading,
next-to-leading and next-to-next-to-leading logarithmic
contributions at TeV energies.  
\\[2mm]
\end{abstract}

\thispagestyle{empty}

\newpage

\section{Introduction}

Experimental and theoretical studies of electroweak interactions have
traditionally explored the range from very low energies, e.g. through
parity violation in atoms, up to energies comparable to the masses of the
$W$- and $Z$-bosons, e.g. at LEP or the Tevatron. The advent of
multi-TeV-colliders like the LHC or a future linear electron-positron
collider during the present decade will give access to a completely new
energy domain.

Once the characteristic energies $s$ are far larger than the masses of the
$W$- and $Z$-bosons, $M_{W,Z}$, multiple soft and collinear
gauge boson emission is kinematically possible. Conversely, exclusive
reactions like electron-positron (or quark-antiquark) annihilation into a
pair of fermions or gauge bosons will receive large negative corrections
from virtual gauge boson emission.
These double logarithmic ``Sudakov'' corrections \cite{Sud,Jac}
proportional to  powers of
${g^2}\ln^2\bigl({s/ M_{W,Z}^2}\bigr)$
are dominant at high energies and thus have to be controlled in
higher orders to arrive at reliable predictions.

The importance of large logarithmic corrections for electroweak
reactions at high energies in one-loop approximation
which may well amount to ten or even twenty
percent  was noticed already
several years ago \cite{Kur,BecCia}. The need for a resummation of
higher orders of these double-logarithmic terms in the context of
electroweak interactions was first emphasized in~\cite{KuhPen} which also
contains a first discussion of this resummation. In particular,
it was shown  that the double logarithms do not depend on the
details of the mechanism of the gauge boson mass generation.
The issue is complicated by the appearance of massive ($W$, $Z$) and
massless ($\gamma$) gauge bosons in the $SU_L(2)\times U(1)$ theory, which
necessarily have to be treated on a different footing.
A complete analysis of this problem in the double
or leading logarithmic (LL) approximation  by the systematic separation
of soft ($\omega_\gamma \le M$) and hard ($\omega_\gamma \ge M$) photons
was given in~\cite{Fad}. In two loops the results of this
approach  essentially based  on the concept of
infrared evolution equations have been confirmed by explicit
calculations in~\cite{HKK}.

The large  coefficient in front of the single logarithmic term in the one-loop
corrections to the  electroweak  amplitudes (see, e.g.~\cite{PasVel})
suggests that subleading terms play an
important role also in higher orders, as long as realistic energies
of order TeV are under consideration.
Motivated by this observation a systematic evaluation of the
next-to-leading logarithmic  (NLL) terms
for the neutral current massless four fermion process was performed in
ref.~\cite{KPS}. Indeed one finds sizeable two-loop effects both for the total cross
section, for the left-right and for the forward-backward asymmetry. A
subclass of NLL corrections for general electroweak processes was
subsequently  evaluated in~\cite{Mel1} without,  however, the very
important {\em angular dependent} contributions.

Various authors have also extracted the  double and
single logarithmic corrections from the complete one-loop
calculations \cite{Bec,DenPoz}.
The analysis for the general electroweak
processes given in~\cite{DenPoz} is in full agreement
with~\cite{KPS}, whereas a different prescription
to separate the QED contribution is adopted in~\cite{Bec}.
Higher order heavy fermion mass effects on the asymptotic high energy
behavior of the electroweak amplitudes were discussed in~\cite{Mel2}.
The incomplete cancellation of the real and virtual electroweak
double logarithmic
corrections in the inclusive cross sections
was investigated in~\cite{CCC}.

Following the approach of \cite{KPS} which in turn is based on the
investigations in the context of QCD
\cite{CorTik,FreTay,APV,MueCol,Sen1,Sen2,Bot,Mag,SterLaen}, this paper is devoted to
the derivation of the next-to-next-to-leading logarithmic
(NNLL) terms for the massless neutral current four-fermion
cross sections. In Section~2 we present as a first step the NNLL form
factor which describes the scattering amplitude in an external Abelian
field for the  $SU(N)$ gauge theory.
The derivation is based on the evolution equation derived in
ref.~\cite{MueCol,Sen1,Sen2}. 
In Section~3 we then 
generalize the result to the four fermion process in the
$SU(N)$ gauge theory. After factoring off the collinear logarithms we
use an evolution equation for the remaining amplitudes which is governed
by an angular dependent soft anomalous dimension matrix 
\cite{Sen2,Bot,SterLaen}. 

Finally we apply this result to  electroweak processes in Section~4.
To identify the pure
QED infrared logarithms which are compensated by soft real photon radiation
we combine the hard evolution equation which governs the
dependence of the amplitudes on $s$ with the infrared evolution equation
\cite{Fad}. The latter describes the dependence of the amplitude
on a fictitious photon mass which serves as an infrared regulator
and drops out after including the effect of
the soft photon emission. The hard and infrared
evolutions are matched by fixing the initial conditions at the scale
$M_{W,Z}$.
Section~4 also contains a discussion of the numerical implications of
our result. A brief summary and conclusions are given in Section~5.

\section{The  Abelian form factor in the Sudakov limit}

Let us first consider
the vector form factor  which determines the
fermion scattering amplitude in an external Abelian field for the
$SU(N)$ gauge model.  In the Born approximation,
\be
{\cal F}_B=\bar\psi(p_2)\gm_\mu\psi(p_1)\, ,
\ee
where
$p_1$ denotes the incoming and $p_2$  the outgoing momentum.

There are two ``standard'' regimes of the Sudakov limit
$s=(p_1-p_2)^2 \to-\infty$: (i)
on-shell massless fermions, $p_1^2=p_2^2=0$ and
gauge bosons with a small non-zero mass $M^2\ll Q^2$
\cite{Jac}, or (ii) slightly off-shell fermions $p_1^2=p_2^2=-M^2$,
and massless gauge boson \cite{Sud}.
Let us consider the first case and choose, for convenience,
$p_{1,2} = (Q/2,0,0,\mp Q/2)$ so that  $2 p_1 p_2 = Q^2=-s$.
The asymptotic $Q$-dependence of the form factor in  this limit
is governed by the evolution equation  \cite{MueCol,Sen1}
\be
{\partial\over\partial\ln{Q^2}}{\cal F}=
\left[\int_{M^2}^{Q^2}{\dd x\over x}\gm(\al(x))+\zeta(\al(Q^2))
+\xi(\al(M^2)) \right] {\cal F} \, .
\label{evoleqf}
\ee
Its solution is
\be
{\cal F}=F_0(\al(M^2))\exp \left\{\int_{M^2}^{Q^2}{\dd x\over x}
\left[\int_{M^2}^{x}{\dd x'\over x'}\gm(\al(x'))+\zeta(\al(x))
+\xi(\al(M^2))\right]\right\}{\cal F}_B \, .
\label{evolsolf}
\ee
The  LL approximation includes all the terms of the
form $\al^n\log^{2n}(Q^2/M^2)$ and is determined by the
one-loop value of  $\gm(\al)$.
The  NLL approximation includes all the terms of the
form $\al^n\log^{2n-m}(Q^2/M^2)$ with $m=0,~1$.
This requires the one-loop values of $\gm(\al)$ and $\zeta(\al)+\xi(\al)$
and using the one-loop running of $\al$ in $\gm(\al)$.
The  NNLL approximation includes all the terms of the
form $\al^n\log^{2n-m}(Q^2/M^2)$ with $m=0,~1,~2$.
In this case
$\gm(\al)$ is required up to ${\cal O}(\al^2)$,
$\zeta(\al)$, $\xi(\al)$ and $F_0(\al)$
up to ${\cal O}(\al)$ together with the two-loop running of $\al$
in $\gm(\al)$ and one-loop running of $\al$
in $\zeta(\al)$.

The functions entering  the evolution equation can 
be determined   by comparing eq.~(\ref{evolsolf}) 
expanded  in the  coupling constant to the  asymptotic, 
{\em i.e.} leading in $M^2/Q^2$,  fixed order result for the 
form factor.  To compute this fixed order asymptotic result
we apply the expansion by regions approach formulated in~\cite{BenSmi} and
discussed using characteristic two-loop examples in~\cite{SmiRak}.
It consists of the following steps:
(i) consider various regions of a loop momentum $k$ and expand, in
every region, the integrand in Taylor series with respect to the
parameters that are there considered small,
(ii) integrate the  expanded integrand
over the whole integration domain of the loop momenta,
(iii) put to zero any scaleless integral.
In step (ii) dimensional regularization  
\cite{dimreg} with $d=4-2\ep$ space-time
dimensions  is used to handle the divergences.
The following regions are relevant in the considered version (i) of
the Sudakov limit \cite{Ste}:
\bea
\label{h}
\mbox{{\em hard} (h):} && k\sim Q\, ,
\nn \\
\label{1c}
\mbox{{\em 1-collinear} (1c):} && k_+\sim Q,\,\,k_-\sim M^2/Q\, ,
\,\, \uk \sim M\,,
\nn \\
\label{2c}
\mbox{{\em 2-collinear} (2c):} && k_-\sim Q,\,\,k_+\sim M^2/Q\, ,
\,\,\uk \sim M \, ,
\nn \\
\label{soft}
\mbox{{\em soft} (s):} && k\sim M \, .
\label{us}
\eea
Here $k_{\pm} =k_0\pm k_3, \, \uk=(k_1,k_2)$. By $k\sim Q$, etc.
we mean that any component of $k_{\mu}$ is of order $Q$.
In one loop  this leads to
the following decomposition \cite{KPS}
\be
{\cal F}^{(1)}=\left(\Delta_h+\Delta_c+\Delta_s\right){\cal F}_B\, ,
\label{1loopfreg}
\ee
\bea
\Delta^{(1)}_h&=&C_F\left(-{2\over\ep^2}+{1\over\ep}\left(2\ln(Q^2)-3\right)
-\ln^2(Q^2)+3\ln(Q^2)+{\pi^2\over 6}-8\right)  \, ,
\nn \\
\Delta^{(1)}_c&=&C_F\left({2\over\ep^2}-{1\over\ep}\left(2\ln(Q^2)-4\right)
+2\ln(Q^2)\ln(M^2)-\ln^2(M^2)-4\ln(M^2)-{5\pi^2\over 6}+4\right)\,,
\nn \\
\Delta^{(1)}_s&=&C_F\left(-{1\over\ep}+\ln(M^2)+{1\over 2}\right)\,.
\label{1loopdel}
\eea
where  $C_F=(N^2-1)/(2N)$ is the  quadratic Casimir operator of
the fundamental representation of the $SU(N)$ group and the 
subscript $c$ denotes the contribution of both collinear regions.
The 't~Hooft scale $\mu$ has been dropped
in the argument of the logarithms as well as the factor
$(4\pi e^{-\gm_E\ep}(\mu^2))^{\ep}$ per loop.
For a perturbative function $f(\al)$ we define
\be
f(\al)=\sum_n\left(\al\over 4\pi\right)^nf^{(n)}\, .
\ee
The contribution of all the regions add up to the well known
result
\be
{\cal F}^{(1)}=-C_F\left(\ln^2\left({Q^2\over M^2}\right)
-3\ln\left({Q^2\over M^2}\right)
+{7\over 2}+{2\pi^2\over 3}\right){\cal F}_B\, .
\label{1loopftr}
\ee
On the other hand the one-loop form factor can be written as
\be
{\cal F}^{(1)}=\left({1\over 2}\gm^{(1)}\ln^2\left({Q^2\over M^2}\right)
+\left(\xi^{(1)}+\zeta^{(1)}\right)\ln\left({Q^2\over M^2}\right)
+F_0^{(1)}\right){\cal F}_B\, .
\label{1loopf}
\ee
The expansion by regions is very efficient for determination
of the parameters of the evolution equation.  Indeed, 
it not necessary to compute the complete asymptotic result in order  
to obtain the functions parameterizing the logarithmic
contributions. In the process of the scale separation 
through the  expansion by regions 
the logarithmic contributions show up as the singularities
of the contributions from different regions which cancel in the 
total result.  Thus one can  identify the regions relevant 
for determining  a given parameter of the evolution equation 
and compute them  separately up to the required accuracy.
For example, the anomalous dimensions $\gamma(\alpha)$ and  
$\zeta(\alpha)$  are known to be independent
on the infrared cutoff and are completely determined by the 
contribution from the hard loop momentum \cite{MueCol,Sen1}. 
If dimensional regularization
is used for the infrared divergences of the hard loop momentum contribution,
as in our approach, the anomalous dimensions $\gm(\al)$ and $\zeta(\al)$
are given  by the coefficients of the double and
single poles of the hard  contribution to the 
exponent~(\ref{evolsolf}), respectively \cite{Sen1,Mag}. 
On the contrary, the functions
$\xi(\al)$ and  $F_0(\al)$ fix the initial conditions
for the evolution equation. They are not universal  
and depend on the infrared sector of the model. 
Furthermore,  the  values of  $\xi(\al)$ and  $F_0(\al)$ depend  
on the definition of the lower integration limits in eq.~(\ref{evolsolf}).  
To determine  the function $\xi(\al)$ one has to know
also the singularities of the collinear region contribution
while $F_0(\al)$  requires the complete information on the contributions
of  all the regions.

From the first line of eq.~(\ref{1loopdel})  we find
the one-loop anomalous dimensions
\bea
\gm^{(1)}&=&-2C_F \,,
\nn \\
\zeta^{(1)}&=&3C_F\,.
\label{1loop1}
\eea
With the above values of $\gamma^{(1)}$ and  
$\zeta^{(1)}$ it is straightforward 
to obtain the one-loop result for the remaining functions  
\bea
\xi^{(1)} &=&0 \,,
\\
F_0^{(1)}&=& -C_F
\left({7\over 2}+{2\pi^2\over 3}\right) \,
\nn
\label{1loop2}
\eea
by comparing eqs.~(\ref{1loopf}) and~(\ref{1loopftr}).
Note that in the Born approximation  $F_0^{(0)}=1$.

A similar decomposition can be performed in two loops
\be
{\cal F}^{(2)}=\left(\Delta_{hh}+\Delta_{hc}+\Delta_{cc}+\ldots\right)
{\cal F}_B\, ,
\label{2loopf}
\ee
Only the hard-hard part is now available \cite{MMN}. 
However, this information is sufficient to determine 
the two-loop value $\gamma^{(2)}$.
Beside the running of the coupling constant, $\gamma^{(2)}$ is 
the only two-loop quantity we need for the NNLL approximation.
It reads  \cite{KodTre}
\bea
\gm^{(2)}&=&-2C_F\left[\left({67\over 9}-{\pi^2\over 3}\right)C_A
-{20\over 9}T_Fn_f\right] \,,
\label{2loop}
\eea
for $\al$ defined in the $\overline{MS}$ scheme.
Here $C_A=N$ is the  quadratic Casimir operator of
the adjoint  representation, $T_F=1/2$ is the
index of the fundamental representation and $n_f$
is the number of light (Dirac) fermions.

Let us consider the two-loop corrections.
The LL, NLL and NNLL approximations are given respectively by
\bea
{\cal F}^{(2)}_{LL}&=&
{1\over 8}{(\gm^{(1)})^2}\ln^4\left({Q^2\over M^2}\right){\cal F}_B\,,
\label{llf}\\
{\cal F}^{(2)}_{NLL}&=&
{1\over 2}\left(\zeta^{(1)}-{1\over 3}\bt_0\right)
\gm^{(1)}\ln^3\left({Q^2\over M^2}\right){\cal F}_B\,,
\label{nllf}\\
{\cal F}^{(2)}_{NNLL}&=&{1\over 2}\left({\gm^{(2)}+
\left(\zeta^{(1)}-\bt_0\right)\zeta^{(1)}
+F_0^{(1)}\gm^{(1)}}\right)
\ln^2\left({Q^2\over M^2}\right){\cal F}_B\,,
\label{nnllf}
\eea
where $\bt_0=11C_A/3-4T_Fn_f/3$ is the one-loop $\bt$-function
provided the normalization point of $\al$
is $M$. The two-loop running of the
coupling constant  in the leading order $\gm(\al)$
starts to contribute in the three-loop NNLL approximation.

The presence of a scalar particle
in the fundamental representation
with no Yukawa coupling to fermions
leads only to a modification of the
$\gm$- and $\bt$-functions.
One scalar boson with the mass  much less than
$Q$ gives the additional contribution of $16C_FT_F/9$ to $\gm^{(2)}$
and the additional contribution of $-T_F/3$ to $\bt_0$.

For the standard model inspired
case of the  $SU(2)_L$ and $U(1)$ gauge groups  with
$n_f=6$ and one charged scalar  boson either in 
the fundamental representation of $SU(2)$ or of the unit
$U(1)$ charge up to NNLL approximation we have
\bea
{\cal F}^{(1)}&=&\left[-{3\over 4}\ln^2\left({Q^2\over M^2}\right)
+{9\over 4}\ln\left({Q^2\over M^2}\right)
-\left({21\over 8}+{\pi^2\over 2}\right)\right]{\cal F}_B\,,
\nn \\
{\cal F}^{(2)}&=&\left[{9\over 32}\ln^4\left({Q^2\over M^2}\right)
-{19\over 48}\ln^3\left({Q^2\over M^2}\right)
-\left({463\over 48}-{7\pi^2\over 8}\right)\ln^2\left({Q^2\over M^2}\right)
\right]{\cal F}_B\,,
\label{fsu2}
\eea
and
\bea
{\cal F}^{(1)}&=&\left[-\ln^2\left({Q^2\over M^2}\right)
+3\ln\left({Q^2\over M^2}\right)
-\left({7\over 2}+{2\pi^2\over 3}\right)\right]{\cal F}_B\, ,
\nn\\
{\cal F}^{(2)}&=&\left[{1\over 2}\ln^4\left({Q^2\over M^2}\right)
-{52\over 9}\ln^3\left({Q^2\over M^2}\right)
+\left({625\over 18}+{2\pi^2\over 3}\right)\ln^2\left({Q^2\over M^2}\right)
\right]{\cal F}_B\,,
\label{fu1}
\eea
respectively.  The relatively
small coefficient  of the  LL terms and the
large coefficient of the  NNLL terms in the
form factor are clearly indicative
of the  importance of the NNLL corrections and, as we will see,
reflect the general structure of the logarithmically enhanced
electroweak corrections.

\section{The four fermion amplitude}
Let us now investigate the four-fermion scattering
at fixed angles  in the limit when
all the  invariant energy and momentum transfers
of the process are far larger than the
gauge boson mass,  $|s|\sim |t| \sim |u| \gg M^2$.
The analysis of the four
fermion amplitude is  complicated  by the extra kinematical variable and
the presence of different
``color'' and Lorentz structures.
We adopt the following notation
\bea
{\cal A}^\lm&=&
\bar{\psi_2}t^a\gm_\mu{\psi_1}
\bar{\psi_4}t^a\gm_\mu{\psi_3}\, , \nn \\
{\cal A}^d_5&=&
\bar{\psi_2}\gm_\mu\gm_5{\psi_1}
\bar{\psi_4}\gm_\mu\gm_5{\psi_3}\, , \nn \\
{\cal A}^\lm_{LL}&=&
\bar{\psi_2}_Lt^a\gm_\mu{\psi_1}_L
\bar{\psi_4}_Lt^a\gm_\mu{\psi_3}_L\, ,  
\label{basis}\\
{\cal A}_{LR}^d&=&
\bar{\psi_2}_L\gm_\mu{\psi_1}_L
\bar{\psi_4}_R\gm_\mu{\psi_3}_R\,, \nn
\eea
etc.
Here $t^a$ denotes the $SU(N)$ generator, $p_i$  the momentum of the
$i$th fermion and
$p_1$, $p_3$ are incoming, and $p_2$, $p_4$ outgoing momenta
respectively.
Hence $t=(p_1-p_4)^2=-sx_-$ and $u=(p_1+p_3)^2=-sx_+$
where $x_{\pm}=(1\pm\cos\theta)/2$
and $\theta$ is the angle between the spatial
components of ${p}_1$ and  ${p}_4$.
The complete  basis  consists
of four independent  chiral amplitudes, each of them
of two possible color structure. For the moment we
consider a parity conserving theory, hence
only two chiral   amplitudes are not degenerate.
The Born amplitude is given by
\be
{\cal A}_{B}={ig^2\over s}{\cal A}^\lm\,.
\label{borna}
\ee
The collinear  divergences in the hard part of the
virtual corrections and the corresponding ``collinear''
logarithms are known to
factorize. They are
responsible, in particular, for the double logarithmic
contribution and depend only on the
properties of the external on-shell particles
but not on  a specific process
\cite{CorTik,FreTay,APV,MueCol,Sen1,Sen2}.
This fact is especially clear if a physical (Coulomb or axial)
gauge is used for the calculation. In this gauge the collinear divergences
are present only in the self energy insertions to the
external particles \cite{FreTay,Sen1,Sen2}.
Thus, for each fermion-antifermion pair
of the four-fermion amplitude the collinear logarithms
are the same as for the form factor ${\cal F}$ discussed in the
previous section.   Let us denote by
$\tilde {\cal A}$   the amplitude with
the collinear logarithms  factored out. For convenience
we separate from  $\tilde {\cal A}$  all the corrections entering
eq.~(\ref{evolsolf}) so that
\be
{\cal A}={ig^2\over s}\left({{\cal F}\over {\cal F}_B}\right)^2\tilde{\cal A}\,.
\ee
The resulting amplitude $\tilde {\cal A}$ contains
the logarithms of the ``soft''  nature  corresponding to
the soft divergences of the hard region contribution 
and the renormalization group   logarithms.
It can be  represented as a vector
in the color/chiral basis
and satisfies the following evolution equation 
\cite{Sen2,Bot,SterLaen}:
\be
{\partial \over \partial \ln{Q^2}}\tilde {\cal A}=
{\bf \chi}(\al(Q^2))\tilde {\cal A} \, ,
\label{evoleqa}
\ee
where  $\chi(\al)$ is the matrix of the soft
anomalous dimensions. 
Note that we do not include to eq.~(\ref{evoleqa})
the  pure   renormalization group   logarithms
which can be absorbed by fixing the normalization scale
of $g$ in the Born amplitude~(\ref{borna}) to be $Q$.
The solution of eq.~(\ref{evoleqa}) reads
\be
\tilde {\cal A}=
\sum_i\tilde {\cal A}_{0i}(\al(M^2))\exp{\left[\int_{M^2}^{Q^2}
{\dd x\over x}\chi_i(\al(x))\right]}\,,
\label{evolsola}
\ee
where $\chi_i(\al)$ are eigenvalues of $\chi(\al)$ and
$\tilde {\cal A}_{0i}(\al)$ 
are $Q$-independent  eigenvectors of $\chi(\al)$
which determine
the initial conditions for the evolution equation at $Q=M$.
Similar to the function $F_0(\al)$ they get contributions from all the regions
while the matrix of the soft
anomalous dimensions is given by 
the coefficients of the  single pole of the hard region 
contribution to the exponent~(\ref{evolsola}) 
\cite{KPS,Sen2}.
Strictly speaking the matrices $\chi(\al(Q^2))$ for different values of $Q$
do not commute and the solution is given by
the path-ordered exponent \cite{Sen2}.
The NLL approximation is given by the
one-loop value of $\chi(\al)$ while the NNLL approximation requires
$\tilde {\cal A}_{0i}(\al)$ up to ${\cal O}(\al)$ together with
the  one-loop running of $\al$ in $\chi(\al)$.

In one loop the elements of the matrix $\chi(\al)$ do not
depend on chirality and read \cite{KPS}
\bea
\chi^{(1)}_{\lm \lm} &=&
-2C_A\left(\ln\left({x_+}\right)+i\pi\right)+
4\left(C_F-{T_F\over N}\right)\ln\left({x_+\over x_-}\right)\, , \nn \\
\chi^{(1)}_{\lm d} &=&4{C_FT_F\over N}\ln\left({x_+\over x_-}\right)\, , \nn \\
\chi^{(1)}_{d \lm} &=& 4\ln\left({x_+\over x_-}\right)\, , \label{china} \\
\chi^{(1)}_{d d} &=& 0\, .  \nn
\eea
In higher
orders the matrix $\chi(\al)$ may be non-degenerate for the different
chiral components of the basis.
In the Abelian case, there are no different color
amplitudes and there is only one anomalous dimension
\be
\chi^{(1)}=4\ln\left({x_+\over x_-}\right)\, .
\label{chia}
\ee

In terms of the functions introduced above
the one-loop correction reads
\bea
{\cal A}^{(1)}&=&{ig^2\over s}
\left[\left(\gm^{(1)}\ln^2\left({Q^2\over M^2}\right)+
\left(2\xi^{(1)}+2\zeta^{(1)}+\chi^{(1)}_{\lm \lm}\right)
\ln\left({Q^2\over M^2}\right)
+2F_0^{(1)}\right){\cal A}^\lm\right.
\nn\\
&+&\left.\chi^{(1)}_{\lm d}
\ln\left({Q^2\over M^2}\right){\cal A}^d+\tilde {\cal A}_0^{(1)}\right]\,,
\label{1loopa}
\eea
where
$\tilde {\cal A}_0^{(1)}=\sum_i\tilde {\cal A}^{(1)}_{0i}
=\tilde {\cal A}^{(1)}|_{Q^2=M^2}$
has the following decomposition
\be
\tilde {\cal A}_0^{(1)}=
\tilde A^{(1)}_0{}^\lm_{LL}{\cal A}^\lm_{LL}+
\tilde A^{(1)}_0{}^\lm_{LR}{\cal A}^\lm_{LR}+\ldots\,.
\ee
 For the present two-loop analysis
of the annihilation cross section only
the real part of the coefficients $\tilde A_0^{(1)}$ is needed,
\bea
{\rm Re}\left[\tilde A^{(1)}_0{}^\lm_{LL}\right]&=&
\left(C_F-{T_F\over N}\right)f(x_+,x_-)
+{C_A}\left({85\over 9}+{\pi^2}\right)-{20\over 9}T_Fn_f\,,
\nn \\
{\rm Re}\left[\tilde A^{(1)}_0{}^\lm_{LR}\right]&=&
-\left(C_F-{T_F\over N}-{C_A\over 2}\right)f(x_-,x_+)
+{C_A}\left({85\over 9}+{\pi^2}\right)-{20\over 9}T_Fn_f\,,
\nn \\
{\rm Re}\left[\tilde A^{(1)}_0{}^d_{LL}\right]&=&
{C_FT_F\over N}f(x_+,x_-)\,,
\label{rea0} \\
{\rm Re}\left[\tilde A^{(1)}_0{}^d_{LR}\right]
&=&-{C_FT_F\over N}f(x_-,x_+)\,,
\nn
\eea
where
\be
f(x_+,x_-)={2\over x_+}\ln x_-
+{x_--x_+\over x_+^2}\ln^2x_-\,.
\ee
A scalar particle
in the fundamental representation
with no Yukawa coupling to fermions
gives the
additional contribution of $-8T_F/9$ to the first two lines of
eq.~(\ref{rea0}).

The two-loop correction is obtained by the direct generalization of
the form factor analysis. The only complication
is related to the matrix structure of
eq.~(\ref{evolsola}):
\bea
{\cal A}^{(2)}_{LL}&=&{ig^2\over s}{{(\gm^{(1)})^2}\over 2}
\ln^4\left({Q^2\over M^2}\right){\cal A}^\lm\,,
\label{lla}\\
{\cal A}^{(2)}_{NLL}&=&
{ig^2(Q^2)\over s}\left[\left(2\zeta^{(1)}+\chi^{(1)}_{\lm \lm}
-{1\over 3}\bt_0
\right){\cal A}^\lm+\chi^{(1)}_{\lm d}{\cal A}^d\right]{\gm^{(1)}}
\ln^3\left({Q^2\over M^2}\right)\,,
\label{nlla}\\
{\cal A}^{(2)}_{NNLL}&=&
{ig^2(Q^2)\over s}\left[\left(\gm^{(2)}+
\left(2{\zeta^{(1)}}-\bt_0\right)
\zeta^{(1)}+2F_0^{(1)}\gm^{(1)}+{1\over 2}
\left(\left(4\zeta^{(1)}-\bt_0\right)\chi^{(1)}_{\lm \lm}
+{\chi^{(1)}_{\lm \lm}}^2\right.
\right.\right.
\nn\\
&&\left.\left.\left.+\chi^{(1)}_{d \lm}\chi^{(1)}_{\lm d}\right)\right)
{\cal A}^\lm +{1\over 2}\left(
\left(4\zeta^{(1)}-\bt_0\right)\chi^{(1)}_{\lm d}
+\chi^{(1)}_{\lm d}\chi^{(1)}_{\lm \lm}+
\chi^{(1)}_{\lm d}\chi^{(1)}_{d d}\right){\cal A}^d
+\gm^{(1)}\tilde {\cal A}_0^{(1)}\right]
\nn\\
&&\times\ln^2\left({Q^2\over M^2}\right)\,.
\label{nnlla}
\eea
The structure of the infrared
singularities of the  hard part of the
two-loop corrections presented here is in full agreement with
the result of~\cite{Cat} which was confirmed by 
explicit calculation \cite{AGOT}.

To illustrate the significance
of the subleading  contributions
let us again  discuss the standard model inspired
example  considered in the  previous section.
Having  the result for the amplitudes it is straightforward
to compute the one- and two-loop corrections to the total
cross section  of the four-fermion annihilation process
using the standard formulae.
For the  annihilation process one has to
make the analytical continuation of the above
result to the Minkowskian region of negative $Q^2=-s$
according to $s+i0$ prescription.
Although the above approximation is formally not valid for small angles
$\theta <M/\sqrt{s}$
we can integrate the  differential cross section
over all angles to get a result with the logarithmic accuracy.
In this way we obtain for the case of $SU(2)_L$ group 
\bea
\sigma^{(1)}&=&\left[-3\ln^2\left({s\over M^2}\right)
+{80\over 3}\ln\left({s\over M^2}\right)
-\left({25\over 9}+3\pi^2\right)\right]\sigma_B\,,
\nn \\
\sigma^{(2)}&=&\left[{9\over 2}\ln^4\left({s\over M^2}\right)
-{449\over 6}\ln^3\left({s\over M^2}\right)
+\left({4855\over 18}+{37\pi^2\over 3}\right)\ln^2\left({s\over M^2}\right)
\right]\sigma_B\,,
\label{sigsu2s}
\eea
and
\bea
\sigma^{(1)}&=&\left[-3\ln^2\left({s\over M^2}\right)
+{26\over 3}\ln\left({s\over M^2}\right)
+\left({218\over 9}-3\pi^2\right)\right]\sigma_B\,,
\nn \\
\sigma^{(2)}&=&\left[{9\over 2}\ln^4\left({s\over M^2}\right)
-{125\over 6}\ln^3\left({s\over M^2}\right)
-\left({799\over 9}-{37\pi^2\over 3}\right)\ln^2\left({s\over M^2}\right)
\right]\sigma_B\,,
\label{sigsu2o}
\eea
for the initial and final state fermions of the same
or opposite isospin, respectively.
Here $\sigma_B$ is the Born cross section
with the $\overline{MS}$
couplings constant normalized at the scale $\sqrt{s}$.

For the $U(1)$  group we have
\bea
\sigma^{(1)}&=&\left[-4\ln^2\left({s\over M^2}\right)
+12\ln\left({s\over M^2}\right)
-\left({382\over 9}-{4\pi^2\over 3}\right)\right]\sigma_B\, ,
\nn\\
\sigma^{(2)}&=&\left[8\ln^4\left({s\over M^2}\right)
-{532\over 9}\ln^3\left({s\over M^2}\right)
+\left({1142\over 3}+{16\pi^2\over 3}\right)\ln^2\left({s\over M^2}\right)
\right]\sigma_B\, .
\label{sigu1}
\eea
Similar to the form factor, we observe a relatively
small coefficient  of the  LL terms and a
large coefficient of the  NNLL terms.

\section{NNL  logarithms in electroweak processes
at high energies}
We are interested in the process
$f'\bar f'\rightarrow f\bar f$.
In the Born approximation, its amplitude is of
the following form
\be
A_{B}={ig^2\over s}\sum_{I,J=L,R}\left(T^3_{f'}T^3_{f}+
t^2_W{Y_{f'}Y_{f}\over 4}\right)A^{f'f}_{IJ}\, ,
\label{aborn}
\ee
where
\be
A^{f'f}_{IJ}=\bar f_I'\gamma_\mu f_I'
\bar f_J\gamma_\mu f_J \, ,
\ee
$t_W=\tan{\theta_W}$ with $\theta_W$ being the Weinberg angle
and  $T_f$ $(Y_f)$ is  the isospin (hypercharge)
of the fermion which depends on the fermion chirality.

To analyze the  electroweak correction to the above process
we use the approximation with
the $W$ and $Z$ bosons of the same mass  $M$, the Higgs boson
of the mass $M_H\sim M$
and massless quarks and leptons.
A fictitious photon mass $\lm$ has to be  introduced
to regularize the infrared divergences.
We insert the mass into the gauge boson propagators ``by hand''
to investigate the leading in $s^{-1}$ behavior
of the amplitudes, leaving
aside the Higgs mechanism of the gauge boson
mass generation.
This approach is gauge invariant
as far as power unsuppressed terms are considered.
The NNLL approximation
is not sensitive to the fine details
of the mass generation because
we need only the hard
part of the   potentially dangerous  self-energy
insertion to the gauge boson propagator.
Indeed, the only effect of the virtual
Higgs boson  is the modification
of the functions $\bt_0$,  $\gm^{(2)}$  and $\tilde A^{(1)}_0{}^\lm$.
The function $\gm^{(2)}$ as well as the running of the coupling
constant in $\gamma^{(1)}$ and $\chi^{(1)}$ 
are determined by the singularities of the hard part of the corrections.
On the other hand, the  vacuum polarization of the off-shell
gauge boson in the Born amplitude contributing
to $\tilde A^{(1)}_0{}^\lm$ is infrared safe
and can be computed in the massless approximation,
i.e. in the leading order in  $s^{-1}$ it
receives only the contribution of the hard region.
The only effect of the Higgs mechanism is that we
have two different  masses $M_Z$ and $M_W=\cos\theta_WM_Z$.
Since $\cos\theta_W\sim 1$ we neglect this difference in our calculation.
The correction due to the heavy gauge boson mass splitting
will be discussed at the end of the section.

Let us first consider the equal mass case $\lm=M$, where
we can work in terms of  the fields of
unbroken phase. In the massless quark approximation
the Higgs boson couples only to the gauge field.
Therefore the result of Sects.~2 and~3
for the $SU(2)_L$ gauge group
with the coupling $g$
and the $U(1)$ gauge group with the coupling $t_Wg$
can be directly applied to the electroweak processes.
For the standard electroweak model with one charged Higgs doublet
one has to replace in the above expressions $n_f\to 2N_g+1/4$  for
$SU(2)_L$ $\beta$-function, $T_Fn_f\to 5N_g/3+1/8$
for $U(1)$ $\beta$-function,  $n_f\to 2N_g+2/5$  for
$SU(2)_L$ $\gamma^{(2)}$ and $\tilde A^{(1)}_0$ coefficients,
and  $T_Fn_f\to 5N_g/3+1/5$ for $U(1)$  
$\gamma^{(2)}$ and $\tilde A^{(1)}_0$ coefficients,
with $N_g=3$ being the number of generations.
For example we have
\bea
\gm^{(2)}&=&{10\over 3}N_g-{65\over 3}+{\pi^2}\, ,
\nn \\
\bt_0&=&-{4\over 3}N_g+{43\over 6}\, ,
\label{su2}
\eea
for  $SU(2)_L$ and
\bea
\gm^{(2)}&=&\left({200\over 27}N_g+{8\over 9}\right)t_W^4{Y_f^2\over 4}\, ,
\nn \\
\bt_0&=&-{20\over 9}N_g-{1\over 6}\, ,
\label{u1}
\eea
for  $U(1)$.

The result for the amplitudes
is obtained by projecting on a relevant initial/final
state with the proper assignment of isospin/hypercharge.
For example, the projection of the  basis~(\ref{basis}) on the
states corresponding to the  neutral current processes reads
${\cal A}_{IJ}^\lm\to T^3_fT^3_{f'}{\cal A}_{IJ}^{f'f}$, 
${\cal A}_{IJ}^d\to {\cal A}_{IJ}^{f'f}$.
The only complication in combinatorics is related to the fact that
now we are having different gauge groups for the fermions of
different chirality.
In particular, the double logarithmic approximation 
is given by  the exponential factor
\be
\exp{\left[
-\left(T_f(T_f+1)+t_W^2{Y_f^2\over 4}+
(f\leftrightarrow f')\right)L(Q^2)  \right]} \, ,
\label{ewfac}
\ee
where
\be
L(Q^2)={g^2\over 16\pi^2}\ln^2\left({Q^2\over M^2}\right),
\ee
and $T_f(T_f+1)=C_F$.

The photon is, however, massless and
the corresponding  infrared divergent contributions  should be
accompanied by  the real soft photon radiation
integrated to some resolution energy  $\omega_{res}$
to get an infrared safe cross
section independent on an auxiliary photon mass.
In practice, the  resolution energy
is much less than the $W$ ($Z$) boson mass
and  the massive gauge bosons are supposed
to be  detected as separate particles.
To study the virtual corrections in the limit of the vanishing
photon mass we follow a
general approach of the infrared evolution equations
developed in~\cite{Fad} (see also  references therein).
It is convenient to use the auxiliary photon
mass $\lm$  as a variable of the infrared evolution
equation below the electroweak scale $M$. The dependence of the
virtual corrections on $\lm$ in the limit $\lm\ll M$ is canceled
by the contribution of the real soft photon radiation.
For  $\omega_{res}\ll M$, the soft photon
emission is of the pure QED nature. Therefore, the kernel of
the infrared evolution  equation which governs the $\lm$
dependence of the virtual corrections to the amplitudes is Abelian.
This dependence is  given by the
QED factor ${\cal U}$
which can be directly obtained from  the
general formulae given above:
\bea
{\cal U}&=&U_0(\al_e)\exp{\left\{
-{\alpha_e(\lm^2)\over 4\pi}\left[\left(\left(Q_f^2+Q_{f'}^2
-\left(
{76\over 27}\left(Q_f^2+Q_{f'}^2
\right)+{16\over 9}Q_fQ_{f'}\right){\alpha_e\over \pi}N_g\right)
\right)\right.\right.}
\nn\\
& &\times\ln^2\left({Q^2\over \lm^2}\right)-\left(3\left(Q_f^2+Q_{f'}^2\right)+
4Q_fQ_{f'}\ln\left({x_+\over x_-}\right)\right)
\ln\left({Q^2\over\lm^2}\right)
+{8\over 27}\left(Q_f^2+Q_{f'}^2\right){\alpha_e\over \pi}N_g
\nn \\
& &
\left.\left.
\times\ln^3\left({Q^2\over\lm^2}\right)\right]+{\cal O}(\al_e^3)\right\}\, ,
\label{QED}
\eea
where $\al_e$ is the $\overline{MS}$
QED coupling constant and
we use the following expressions for the QED functions
\bea
\zeta^{(1)}_e&=&3Q_f^2\, ,
\nn\\
\chi^{(1)}_e&=&4Q_{f'}Q_f
\ln\left({x_+\over x_-}\right)\, ,
\nn\\
\bt_0^e&=&-{32\over 9}N_g\, ,
\label{QEDpar}
\\
\gm^{(2)}_e&=&{320\over 27}N_gQ_f^2\, .
\nn
\eea
The expressions for $\bt_0^e$  and $\gm^{(2)}_e$
can be obtained by  substituting
$T_Fn_f\to 8N_g/3$ to the general formulae.
The coefficient $U_0(\al_e)$ in eq.~(\ref{QED})
is a two-component vector  in the chiral basis.

In a full analogy with the renormalization group
all the information on the non-Abelian gauge dynamics above 
the electroweak scale up to power suppressed contributions
is contained in the initial condition for this
Abelian infrared evolution equation at the point
$\lm=M$.  To fix a relevant initial condition for
the evolution in $\lm$ below the electroweak scale
one has  to subtract the QED virtual correction (\ref{QED})
computed with the photon of the  mass $M$
from the complete result with $\lm=M$ \cite{Fad}.
This leads, in particular, to the modification of the function $\gm(\al)$
so  that  the  double logarithmic exponential
factor becomes
\be
\exp{\left[
-\left(T_f(T_f+1)+t_W^2{Y_f^2\over 4}-s_W^2Q_f^2+
(f\leftrightarrow f')\right)
L(Q^2) \right]} \, ,
\label{col}
\ee
where $s_W=\sin{\theta_W}$. In the NNLL approximation
after the subtraction we get
\bea
\gm^{(2)}&=&-2\left[\left(-{20\over 9}N_g+
{130\over 9}-{2\pi^2\over 3}\right)T_f(T_f+1)-
\left({100\over 27}N_g+{4\over 9}\right)
t^4_W{Y^2_{f}\over 4}\right.
\nn \\
& &\left.+{160\over 27}N_gs_W^4Q_f^2\right]\, .
\label{g2sub}
\eea
A similar subtraction should be done for the parameters
$\zeta^{(1)}$ and $\chi^{(1)}$  which take the form \cite{KPS}
\be
\zeta^{(1)}=3\left(T_f(T_f+1)+t_W^2{Y_f^2\over 4}-s_W^2Q_f^2\right)\, ,
\ee
and
\bea
\chi^{(1)}_{\lm \lm} &=&
-4\left(\ln\left({x_+}\right)+i\pi\right)+
\left(t_W^2Y_{f'}Y_f-4s_W^2Q_{f'}Q_f+2\right)\ln\left({x_+\over x_-}
\right)\, ,
\nn \\
\chi^{(1)}_{\lm d} &=&{3\over 4}\ln\left({x_+\over x_-}\right)\, , \nn \\
\chi^{(1)}_{d \lm} &=& 4\ln\left({x_+\over x_-}\right)\, , \\
\chi^{(1)}_{d d} &=& \left(t_W^2Y_{f'}Y_f-4s_W^2Q_{f'}Q_f\right)
\ln\left({x_+\over x_-}\right)\, . \nn
\eea
For $I$ or $J=R$ the matrix $\chi^{(1)}$ is reduced to
\be
\chi^{(1)}=
\left(t_W^2Y_{f'}Y_f-4s_W^2Q_{f'}Q_f\right)\ln\left({x_+\over x_-}\right)\,.
\ee
At the same time we have some freedom in the definition of the
coefficients $F_0(\al)$, $\tilde A_0(\al)$ and $U_0(\al_e)$.
If we use the one-loop normalization condition   ${\cal U}|_{Q^2=M^2}=1$,
then  $U^{(1)}_0=0$ and no QED subtraction
is necessary for $F^{(1)}_0$ and $\tilde A^{(1)}_0$. In this case  the
QED factor ${\cal U}$ is universal and has no matrix structure.
To summarize, we have  two evolution equations
and corresponding  initial conditions. The coefficients
$F_0(\al)$ and $\tilde A_0(\al)$ give the initial condition
for the hard evolution of the amplitudes in  $Q$
at $Q=M$  while the
above subtraction of the QED contribution
gives the initial condition for the
infrared evolution of the amplitudes in
$\lm$ at $\lm=M$.

The result for the $n$-loop correction to the amplitude~(\ref{aborn}) 
can be decomposed as 
\bea
{A}^{(n)} &=&{A}^{(n)}_{LL} + {A}^{(n)}_{NLL} + 
{A}^{(n)}_{NNLL}+\ldots\, .
\eea
Explicit expressions 
for ${A}^{(1)}_{LL}$ and ${A}^{(1)}_{NLL}$  can be found, for example,
in~\cite{KPS}. In the NNLL approximation one has to take into
account also the one-loop constant contribution corresponding to
$F^{(1)}_0$ and $\tilde {\cal A}^{(1)}_0$ terms of eq.~(\ref{1loopa}).
It reads
\bea
{\lefteqn{
a\, {A}^{(1)}_{NNLL}\, =}}  \\
&&{ig^2\over s}\sum_{I,J=L,R} \left\{-\left({7\over 2}+{2\pi^2\over 3}\right)
\left[T_f(T_f+1)+t_W^2{Y_f^2\over 4}+(f\leftrightarrow f')\right]
\left[T^3_{f'}T^3_f+t^2_W{Y_{f'}Y_f\over 4}\right]\right.
\nn \\
&&
+\left(2T^3_{f'}T^3_f+
t^2_W{Y_{f'}Y_f\over 4}\right)t_W^2{Y_{f'}Y_f\over 4}
\bigg[f(x_+,x_-)
\left(\delta_{IR}\delta_{JR}+\delta_{IL}\delta_{JL}\right)
\nn \\
&&
-f(x_-,x_+)\left(\delta_{IR}\delta_{JL}
+\delta_{IL}\delta_{JR}\right)\bigg]
-\left[
\left({20\over 9}N_g+{4\over 9}\right)T^3_{f'}T^3_f+
\left({100\over 27}N_g+{4\over 9}\right)
t^4_W{Y_{f'}Y_f\over 4}\right]
\nn \\
&&
+\left.\left[{1\over 2}f(x_+,x_-)+{170\over 9}+2\pi^2
\right]T^3_{f'}T^3_f
+{3\over 16}f(x_+,x_-)\delta_{IL}\delta_{JL}\right\}
a {A}^{f'f}_{IJ} \, ,
\nn
\eea
where $a = g^2/16\pi^2$ and we keep only the real part of
$\tilde {\cal A}^{(1)}_0$.  

Let us consider the two-loop corrections.
The two-loop LL corrections
to the chiral amplitudes  were obtained in~\cite{Fad}
\bea
{\lefteqn{
a^2\,{A}^{(2)}_{LL}\, =\, {ig^2\over s}\sum_{I,J=L,R}}}  \\
&&
{1\over 2}\left(T_f(T_f+1)+t_W^2{Y_f^2\over 4}-s_W^2Q_f^2+
(f\leftrightarrow f')\right)^2\left[T^3_{f'}T^3_f+
t^2_W{Y_{f'}Y_f\over 4}\right]
L^2(Q^2) {A}^{f'f}_{IJ}\, ,
\nn
\eea
and the two-loop NLL  corrections  with the exception of the  trivial
corrections proportional to $\bt_0$ can be found in~\cite{KPS}
\bea
a^2\,{A}^{(2)}_{NLL} &=& -\, {ig^2\over s}\sum_{I,J=L,R}
\left[T_f(T_f+1)+t_W^2{Y_f^2\over 4}-s_W^2Q_f^2+
(f\leftrightarrow f')\right] \label{smnll} \\
&&\times\left\{3\left[T_f(T_f+1)+t_W^2{Y_f^2\over 4}-s_W^2Q_f^2+
(f\leftrightarrow f')\right]\left[T^3_{f'}T^3_f+
t^2_W{Y_{f'}Y_f\over 4}\right]
\right. \nn
\\
&&+\left[-4\left(\ln\left({x_+}\right)+i\pi\right)
+\ln\left({x_+\over x_-}\right)\left(2+
t_W^2Y_{f'}Y_f\right)
\right]T^3_{f'}T^3_f
+{3\over 4}\ln\left({x_+\over x_-}\right)\delta_{IL}\delta_{JL}
\nn \\
&&\left.
+\ln\left({x_+\over x_-}\right)\left[t_W^2Y_{f'}Y_f-4s_W^2Q_{f'}Q_f
\right]\left[T^3_{f'}T^3_f+
t^2_W{Y_{f'}Y_f\over 4}\right]\right\}L(Q^2)l(Q^2)
{A}^{f'f}_{IJ}\nn \, ,
\eea
where
\be
l(Q^2)={g^2\over 16\pi^2}\ln\left({Q^2\over M^2}\right),
\ee
with $\delta_{IL}=1$ for $I=L$ and zero otherwise.
The second line of  eq.~(\ref{smnll}) corresponds to the
$\zeta^{(1)}$ term of eq.~(\ref{nlla})
while the third and forth lines correspond
to the $\chi^{(1)}$ terms in eq.~(\ref{nlla}).
A part of the $\bt_0$ NLL corrections
is absorbed by choosing the normalization point of
the coupling constants in eq.~(\ref{aborn}) to be $Q$.
The rest is due to the running of the coupling constant
in the double logarithmic integral and corresponds
to the $\bt_0$  term in  eq.~(\ref{nlla}).
It is of the form
\bea
{\lefteqn{
a^2\,{A}^{(2)}_{NLL}|_{\beta_0} \,=\, -\, {ig^2\over
  s}\sum_{I,J=L,R}}} \label{bt0corr} \\
&&{1\over 3}\left[\left({4\over 3}N_g-{43\over 6}\right)T_f(T_f+1)
+\left({20\over 9}N_g+{1\over 6}\right)
t_W^4{Y_f^2\over 4}-{32\over 9}N_gs_W^4Q_f^2+
(f\leftrightarrow f')\right]
\nn\\
&&\times\left[T^3_{f'}T^3_f+
t^2_W{Y_{f'Y_f}\over 4}\right]l(Q^2)L(Q^2)
{A}^{f'f}_{IJ}\, ,
\nn
\eea
provided  the normalization point of the coupling
constants  is $M$ with the exception of the coupling constants
entering the  Born amplitude~(\ref{aborn})
normalized at  the scale $Q$.

Let us consider the NNLL contribution.
For convenience we split it in four parts
\bea
{A}^{(2)}_{NNLL} &=& \Delta_1{A}^{(2)}_{NNLL} + \Delta_2{A}^{(2)}_{NNLL} + 
\Delta_3{A}^{(2)}_{NNLL}+ 
\Delta_4{A}^{(2)}_{NNLL}\, ,
\eea
where the trivial $\bt_0^2$ renormalization group logarithms
which can be absorbed into the running of the coupling constants
in  eq.~(\ref{aborn}) are not included.
The correction  $\Delta_1{A}^{(2)}_{NNLL}$
corresponding to the $\gamma^{(2)}$, 
${\zeta^{(1)}}^2$, $\bt_0\zeta^{(1)}$ and $F_0^{(1)}\gamma^{(1)}$ terms of
eq.~(\ref{nnlla}) is
\bea
{\lefteqn{
a^2\,\Delta_1{A}^{(2)}_{NNLL} \,=\, {ig^2\over
  s}\sum_{I,J=L,R}}} \\
\hspace*{-5mm}&&\left\{-\left[\left(-{20\over 9}N_g+
{130\over 9}-{2\pi^2\over 3}\right)T_f(T_f+1)-
\left({100\over 27}N_g+{4\over 9}\right)
t^4_W{Y^2_{f}\over 4}+
{160\over 27}N_gs_W^4Q_f^2\right.\right.
\nn \\
&&
+(f\leftrightarrow f')\bigg]
+{9\over 2}\left[T_f(T_f+1)+t_W^2{Y_f^2\over 4}-s_W^2Q_f^2
+(f\leftrightarrow f')\right]^2
\nn \\
&&+{3\over 2}\left[\left({4\over 3}N_g-{43\over 6}
\right)T_f(T_f+1)+
\left({20\over 9}N_g+{1\over 6}\right)
t^4_W{Y^2_{f}\over 4}-
{32\over 9}N_gs_W^4Q_f^2+
(f\leftrightarrow f')\right]
\nn \\
&&\left.+\left({7\over 2}+{2\pi^2\over 3}\right)
\left[T_f(T_f+1)+t_W^2{Y_f^2\over 4}-s_W^2Q_f^2+
(f\leftrightarrow f')\right]\left[T_f(T_f+1)+t_W^2{Y_f^2\over 4}
\right.\right.
\nn \\
&&
+(f\leftrightarrow f')\bigg]\bigg\}\left[T^3_{f'}T^3_f+
t^2_W{Y_{f'}Y_f\over 4}\right]l^2(Q^2){A}^{f'f}_{IJ} \, .
\nn
\eea
The correction $\Delta_2{A}^{(2)}_{NNLL}$
corresponding to the $\zeta^{(1)}\chi^{(1)}$
and  $\bt_0\chi^{(1)}$ terms of
eq.~(\ref{nnlla}) reads
\bea
{\lefteqn{
a^2\,\Delta_2{A}^{(2)}_{NNLL} \,=\, {ig^2\over
  s}\sum_{I,J=L,R}}} \\
&&\left\{\left[\left(-4\left(\ln\left({x_+}\right)+i\pi\right)
+\ln\left({x_+\over x_-}\right)\left(2+t_W^2Y_{f'}Y_f\right)
\right)T^3_{f'}T^3_f
+{3\over 4}\ln\left(x_+\over x_-{}\right)\delta_{IL}\delta_{JL}\right]
\right.
\nn \\
&&
\times\left[\left({2\over 3}N_g-{43\over 12}\right)+3
\left(T_f(T_f+1)+t_W^2{Y_f^2\over 4}-s_W^2Q_f^2+
(f\leftrightarrow f')\right)\right]
\nn \\
&&
+4\ln\left({x_+\over x_-}\right)
\left[\left(
\left({10\over 9}N_g+{1\over 12}\right)t_W^4{Y_{f'}Y_f\over 4}
-{16\over 9}N_gs_W^4Q_{f'}Q_f\right)+3
\left(t_W^2{Y_{f'}Y_f\over 4}
-s_W^2Q_{f'}Q_f\right)\right.
\nn \\
&&\left.\left.\times
\left(T_f(T_f+1)+t_W^2{Y_f^2\over 4}-s_W^2Q_f^2+
(f\leftrightarrow f')\right)\right]\left[T^3_{f'}T^3_f+
t^2_W{Y_{f'}Y_f\over 4}\right]\right\}l^2(Q^2)
{A}^{f'f}_{IJ}\nn \, .
\eea
The correction  $\Delta_3{A}^{(2)}_{NNLL}$ 
corresponding to the  ${\chi^{(1)}}^2$
terms of eq.~(\ref{nnlla}) reads
\bea
{\lefteqn{
a^2\,\Delta_3{A}^{(2)}_{NNLL} \,=\, {ig^2\over
  s}\sum_{I,J=L,R}}} \\
&&{1\over 2}\left\{\left(-4\left(\ln\left({x_+}\right)+i\pi\right)
+\ln\left({x_+\over x_-}\right)
\left(2+t^2_WY_{f'}Y_f-4s_W^2Q_{f'}Q_f\right)\right)^2T^3_{f'}T^3_f
+\ln\left({x_+\over x_-}\right)\right.
\nn
\\
&&
\times
\left[-4\left(\ln\left({x_+}\right)+i\pi\right)
+2\ln\left({x_+\over x_-}\right)
\left(1+t^2_WY_{f'}Y_f-4s_W^2Q_{f'}Q_f\right)\right]
\nn \\
&&
\times\left[{3\over 4}\delta_{IL}
\delta_{JL}+t_W^2Y_{f'}Y_fT^3_{f'}T^3_f\right]
+\ln^2\left({x_+\over x_-}\right)\left[3\left(T^3_{f'}T^3_f+
t^2_W
{Y_{f'}Y_f\over 4}\delta_{IL}\delta_{JL}\right)
\right.
\nn \\
&&\left.\left.+\left(t_W^2Y_{f'}Y_f-4s_W^2Q_{f'}Q_f
\right)^2
t^2_W{Y_{f'}Y_f\over 4}\right]\right\}
 l^2(Q^2){A}^{f'f}_{IJ}  \, .
\nn
\eea
The correction  $\Delta_4{A}^{(2)}_{NNLL}$
corresponding to the real part of the $\gm^{(1)}\tilde {\cal A}^{(1)}_0$
term of eq.~(\ref{nnlla}) reads
\bea
{\lefteqn{
a^2\,\Delta_4{A}^{(2)}_{NNLL} \,=\, -\, {ig^2\over
  s}\sum_{I,J=L,R}}} \\
&&\left(T_f(T_f+1)+t_W^2{Y_f^2\over 4}-s_W^2Q_f^2+
(f\leftrightarrow f')\right)\left\{\left(2T^3_{f'}T^3_f+
t^2_W{Y_{f'}Y_f\over 4}\right)t_W^2{Y_{f'}Y_f\over 4}
\right.
\nn \\
&&
\times\bigg[f(x_+,x_-)
\left(\delta_{IR}\delta_{JR}+\delta_{IL}\delta_{JL}\right)-
f(x_-,x_+)\left(\delta_{IR}\delta_{JL}
+\delta_{IL}\delta_{JR}\right)\bigg]
\nn \\
&&-\left[
\left({20\over 9}N_g+{4\over 9}\right)T^3_{f'}T^3_f+
\left({100\over 27}N_g+{4\over 9}\right)
t_W^4{Y_{f'}Y_f\over 4}\right]
\nn \\
&&
+\left.\left[{1\over 2}f(x_+,x_-)+{170\over 9}+2\pi^2
\right]T^3_{f'}T^3_f
+{3\over 16}f(x_+,x_-)\delta_{IL}\delta_{JL}\right\}l^2(Q^2)
{A}^{f'f}_{IJ}\nn \, .
\eea
With the expression for the  chiral amplitudes
at hand, we can compute the leading and subleading logarithmic corrections
to the basic observables for $e^+e^-\rightarrow f \bar f$.

In the NNLL approximation
one has to take into account also the effect of analytical
continuation to the physical positive real value of the invariant $s$.
For the annihilation processes it is more natural
to normalize  the QED factor at the Minkowskian point $s=M^2$ to 
${\cal U}|_{s=M^2}=1$ so that after the expansion in $\al_e$
it reads
\bea
{\cal U}&=&
\left\{1-{\alpha_e(\lm^2)\over 4\pi}\left[\left(Q_f^2+Q_{f'}^2\right)
\ln^2\left({s\over \lm^2}\right)-\left((3+2i\pi)\left(Q_f^2+Q_{f'}^2\right)+
4Q_fQ_{f'}\ln\left({x_+\over x_-}\right)\right)
\right.\right.
\nn \\
& &
\left.\times\ln\left({s\over\lm^2}\right)
\right]+{\cal O}(\al_e^2)\bigg\}\, ,
\label{expu}
\eea

Let us  consider  the total cross sections
of the quark-antiquark/$\mu^+\mu^-$ production in
the $e^+e^-$ annihilation.
The  LL, NLL and NNLL
corrections to the cross sections to one and two loops read   
\bea
R_{Q\bar Q}\hspace{3mm}
&=&1-1.66\,L(s)+\hspace{2mm}5.31\,l(s)
-\hspace{2mm}8.36\,a+1.93\,L^2(s)-10.59\,L(s)l(s)
+\hspace{2mm}31.40\,l^2(s) \, ,
\nn \\
R_{q\bar q}\hspace{5mm}
&=&1-2.18\,L(s)+20.58\,l(s)-34.02\,a+2.79\,L^2(s)-51.04\,L(s)l(s)+
309.34\,l^2(s)\, ,
\nn\\
R_{\mu^+\mu^-}
&=&1-1.39\,L(s)+10.12\,l(s)-20.61\,a+1.42\,L^2(s)-19.81\,L(s)l(s)+
107.03\,l^2(s)\, ,
\nn\\
 & & \label{finres}
\eea
where  $Q=u,c,t$, $q=d,s,b$,
$R_{Q\bar Q}=\sigma/\sigma_B(e^+e^-\rightarrow Q\bar Q)$ and
so on.
The $\overline{MS}$
couplings in the Born cross section are normalized at $\sqrt{s}$.
Numerically, we have $L(s)=0.07$  $(0.11)$ and $l(s)=0.014$  $(0.017)$
for $\sqrt{s}=1$~TeV and $2$~TeV respectively.
Here $M=M_W$ 
has been chosen for the infrared cutoff and $a = 2.69\cdot 10^{-3}$,
$s_W^2=0.231$ for the $\overline{MS}$ couplings normalized at the gauge boson mass.
The small difference between the two-loop NLL coefficients
in  eq.~(\ref{finres}) and the result of~\cite{KPS}
is due to the $\bt_0$ contribution~(\ref{bt0corr}).

To get the infrared safe result for the semi-inclusive
cross sections  one has to add to the expressions given above
the standard QED corrections due to the soft photon emission
and the {\em pure} QED virtual correction 
which is determined for massless or light fermions of the mass 
$m_f\ll \lm\ll M$ by eqs.~(\ref{QED}),~(\ref{expu}).
To derive  the  QED factor for $\lm$ far less than the
fermion mass $\lm\ll m_f\ll M$ one has to change the kernel of the
infrared evolution equation and match the new solution to eq.~(\ref{QED})
at the point $\lm =m_f$.
The sum of the real and virtual  QED corrections depends
on  $s$, $\omega_{res}$ and  on  the initial/final
fermion masses but not on $M_{Z,W}$.
Note that our analysis implies the resolution energy
for the real photon
emission to be smaller than the heavy boson mass.
If the resolution energy exceeds $M_{Z,W}$ the analysis is more
complicated due to  the fact that the radiation of real photons
is not of Poisson type because of its non-Abelian $SU(2)_L$ component
\cite{Fad}.
In the case of the quark-antiquark final state the strong
interaction also produces the logarithmically
growing terms. They can be read off the results of
Section 1 for the form factor. For massless quarks
the complete ${\cal O}(\al_s^2)$  corrections including
the bremsstrahlung effects can be found in~\cite{MMN}.

 For completeness we give a numerical estimate
of corrections to the  cross section asymmetries.
In the case of  the  forward-backward  asymmetry $A^{FB}$
(the difference of
the cross section averaged over forward and backward semispheres
with respect to the electron beam direction
divided by the total cross section) we get
\bea
R^{FB}_{Q\bar Q}\hspace{3mm}
&=&1-0.09\,L(s)-1.23\,l(s)+\hspace{2mm}1.47\,a+0.12\,L^2(s)
+0.64\,L(s)l(s)-\hspace{2mm}1.40\,l^2(s) \, ,
\nn \\
R^{FB}_{q\bar q}\hspace{3mm}
&=&1-0.14\,L(s)+7.15\,l(s)-10.43\,a+0.02\,L^2(s)-1.31\,L(s)l(s)-
33.46\,l^2(s)\, ,
\nn\\
R^{FB}_{\mu^+\mu^-}
&=&1-0.04\,L(s)+5.49\,l(s)-14.03\,a+0.27\,L^2(s)-6.32\,L(s)l(s)+
21.01\,l^2(s)\, ,
\nn\\
 & & \label{finresfb}
\eea
where $R^{FB}=A^{FB}/A^{FB}_B$.
 For the left-right asymmetry $A^{LR}$ (the difference of the cross sections
of the left and right particles production divided by the
total cross section) we obtain in the same notation
\bea
R^{LR}_{Q\bar Q}\hspace{3mm}
&\hspace*{-4mm}=\hspace*{-4mm}&1-\hspace{2mm}2.34\,L(s)+\hspace{4mm}8.98\,l(s)
-\hspace{4mm}5.73\,a-0.46\,L^2(s)
+\hspace{2mm}7.43\,L(s)l(s)-\hspace{2mm}18.59\,l^2(s) \, ,
\nn \\
R^{LR}_{q\bar q}\hspace{3mm}
&\hspace*{-4mm}=\hspace*{-4mm}&1-\hspace{2mm}1.12\,L(s)+\hspace{2mm}11.86\,l(s)
-\hspace{2mm}15.83\,a-0.81\,L^2(s)
+17.74\,L(s)l(s)-127.05\,l^2(s)\, ,
\nn\\
R^{LR}_{\mu^+\mu^-}
&\hspace*{-4mm}=\hspace*{-4mm}&1-13.24\,L(s)+113.77\,l(s)-139.94\,a-0.79\,L^2(s)+
23.34\,L(s)l(s)-155.36\,l^2(s)\, .
\nn\\
 & \hspace*{-2mm}& \label{finreslr}
\eea
 Finally, for the  left-right asymmetry $\tilde A^{LR}$
(the difference of the cross sections
for the left and right initial state particles divided by the
total cross section) which differs from
$A^{LR}$ for the quark-antiquark final state we have
\bea
\tilde R^{LR}_{Q\bar Q}\hspace{3mm}
&\hspace*{-3mm}=\hspace*{-2mm}&1-\hspace{2mm}2.75\,L(s)+\hspace{2mm}10.07\,l(s)
-\hspace{2mm}9.02\,a-0.91\,L^2(s)
+10.80\,L(s)l(s)-\hspace{2mm}32.10\,l^2(s) \, ,
\nn \\
\tilde R^{LR}_{q\bar q}\hspace{3mm}
&\hspace*{-3mm}=\hspace*{-2mm}&1-\hspace{2mm}1.07\,L(s)+\hspace{2mm}11.56\,l(s)
-\hspace{0mm}15.60\,a-0.77\,L^2(s)
+16.78\,L(s)l(s)-121.56\,l^2(s)\, .
\nn\\
 & & \label{finreslrtil}
\eea
In the $1-2$~TeV region
the two-loop LL, NLL and NNLL corrections to the cross sections
can be as large as  $1-4\%$, $5-10\%$,
and $5-9\%$  respectively.  However, we observe a
{\em significant cancellation}
between  different terms and
the sum of the known two-loop  corrections
amounts of  approximately  $1-2\%$.
The sum of the two-loop  correction to the
asymmetries is even smaller and does not exceed $1\%$ level
with the exception  of the $R^{LR}_{\mu^+\mu^-}$.
For this quantity the relatively large corrections
are the consequence of the  numerically small  Born
approximation.

Let us discuss the accuracy of our result.
At TeV energies the LL, NLL and NNLL corrections 
of eqs.~(\ref{finres})--(\ref{finreslrtil}) provide 
asymptotic expressions for the cross section and the 
asymmetries to one and two loops.
The complete one-loop corrections are known exactly (see \cite{BHM}
for the most general result) and 
we have included the dominant one-loop terms 
in eqs.~(\ref{finres})--(\ref{finreslrtil}) 
to demonstrate the structure of the expansion rather 
than for precise numerical estimates. 
For physical applications, 
the mass difference between the $W$ and the $Z$ gauge boson, 
power suppressed terms and also top quark mass effects can be important.
The effect of the $W$ and $Z$ gauge boson mass difference
on the coefficients of the NLL and NNLL terms
is suppressed as  $(M_Z-M_W)/M\sim 0.1$ while
the leading power corrections can be as large as $M^2/s< 0.01$.
Thus, except for the production of third generation quarks, 
the above expressions approximate
the exact one-loop result with $1\%$ accuracy in the TeV region.
At the same time both effects can be neglected in two-loop approximation.
Therefore, the only essential deviation of the complete
two-loop NLL and NNLL result
from eqs.~(\ref{finres})--(\ref{finreslrtil}) for the
production of the third generation quarks is due to
the large top quark Yukawa coupling. Numerically, 
the corresponding  corrections can be 
as important as the generic non-Yukawa ones.

Finally, let us emphasize that the angular dependent NLL and NNLL terms are
quite important for the cross section and dominate in particular 
the forward-backward asymmetry.

\section{Summary}
In the present paper we employed the evolution
equation approach to analyze
the high energy asymptotic behavior of the four-fermion amplitudes
in the non-Abelian gauge models. The results  were used to
compute the NNLL electroweak corrections to
the  neutral current four-fermion
processes  at high energy
in the massless quark approximation to all orders
in the coupling constants.
We have shown the NNLL approximation to be insensitive to the details
of the gauge boson mass generation as well as to the
Higgs boson mass and self-coupling.

We have calculated the explicit expressions for the one- and two-loop terms which
saturate the NNLL corrections to the basic
observables in the TeV region.
In general, the two-loop
NLL and NNLL corrections exceed the LL contribution in the TeV region
due to the numerically small coefficient in front
of the double logarithmic terms. Hence the truly asymptotic
behavior sets in only at an significantly higher energy.
At the same time the two-loop NNLL corrections are numerically of the 
same magnitude but slightly smaller than the NLL ones, both being in 
the range of $1\%$-$10\%$. 
This could be considered as a signal of  convergence of the logarithmic
expansion at TeV energies. Indeed, the two-loop coefficients in front
of  $\ln^2\bigl(s/M^2\bigr)$ is {(a few units)}$\times\al^2$. 
This is not an unusually large value in a non-Euclidean regime where 
the expansion parameter is $\al$, rather then $\al/(4\pi)$ as 
can be seen in eqs.~(\ref{sigsu2s})--(\ref{sigu1}) 
(see also \cite{Mag,KodTre,AGOT}).
Moreover, we have observed a significant cancellation
between the two-loop LL, NLL and NNLL terms.  
As a result of this cancellation the sum of these two-loop corrections
to the cross sections is of order $1-2\%$ for all the processes.

Thus, if we assume no further growth of the coefficient for the single
logarithmic and non-logarithmic two-loop terms and the observed
pattern of cancellation to hold, we would argue that our NNLL result 
approximates the exact cross sections with $1\%$ accuracy.
The accuracy is less for the production of third generation quarks
where we cannot neglect the large Yukawa coupling to the Higgs boson 
that modifies the NLL and NNLL terms in our formulae.

\vspace{4mm}

{\bf Acknowledgments}\\[3mm]
S.M. would like to thank P. Uwer for fruitful discussions.
The work of J.H.K. and S.M. 
was supported by the DFG under contract FOR~264/2-1, and 
by the BMBF under grant BMBF-05HT1VKA/3.
The work of A.P. was supported by the DFG through grant
KN~365/1-1, by the BMFB through grant 05~HT9GUA~3, by the EC through 
contract ERBFMRX-CT98-0194, and by INTAS through grant 00-00313.
The work of V.S. was supported by the Russian Foundation for Basic
Research through project 01-02-16171, and by INTAS through grant 00-00313.


\begin{thebibliography}{99}
\bibitem{Sud} V.V.~Sudakov, {\em Zh. Eksp. Teor. Fiz.} 30 (1956) 87.

\bibitem{Jac} R.~Jackiw, {\em Ann. Phys.} 48 (1968) 292; 51 (1969) 575.

\bibitem{Kur} M.~Kuroda, G.~Moultaka and D.~Schildknecht, {\em Nucl. Phys.}
              {B350} (1991) 25;\\
              G.~Degrassi and A.~Sirlin, {\em Phys. Rev.} {D46} (1992) 3104.

\bibitem{BecCia} M.~Beccaria {\it et al.}, {\em Phys. Rev.} {D58} (1998)
                 093014;\\
                 P.~Ciafaloni and  D.~Comelli,  {\em Phys. Lett.} {B446}
                 (1999) 278.

\bibitem{KuhPen} J.H.~K\"uhn and A.A.~Penin, Preprint TTP/99--28,
                 hep-ph/9906545.

\bibitem{Fad} V.S.~Fadin, L.N.~Lipatov, A.D.~Martin and M.~Melles,
              {\em Phys. Rev.} D61 (2000)  094002.

\bibitem{HKK}  M.~Hori, H.~Kawamura and J.~Kodaira,
               {\em Phys. Lett.} B491 (2000) 275.

\bibitem{PasVel} G.~Passarino and M.~Veltman,  
                 {\em Nucl. Phys.} B160 (1979) 151.

\bibitem{KPS}  J.H.~K\"uhn, A.A.~Penin and V.A.~Smirnov,
               {\em Eur. Phys. J.}  C17 (2000) 97;
               {\em Nucl. Phys. Proc. Suppl.} 89 (2000) 94.

\bibitem{Mel1}  M.~Melles,  {\em Phys. Rev.} D63 (2001) 034003.

\bibitem{Bec}    M.~Beccaria {\em et al.},  {\em Phys. Rev.} D61 (2000)
                 011301; D61 (2000)  073005;\\
                 M.~Beccaria, F.M.~Renard and C.~Verzegnassi,
                 {\em Phys. Rev.} D63 (2001) 053013.

\bibitem{DenPoz}  A.~Denner and S.~Pozzorini, Preprint PSI-PR-00-15,
                  hep-ph/0010201.

\bibitem{Mel2}  M.~Melles, {\em Phys.Rev.} D64 (2001) 014011.

\bibitem{CCC}  M.~Ciafaloni, P.~Ciafaloni and  D.~Comelli,
               {\em Phys. Rev. Lett.} 84 (2000) 4810;
               {\em Nucl. Phys.} B589 (2000) 359.

\bibitem{CorTik} J.M.~Cornwall and  G.~Tiktopoulos, {\em Phys. Rev. Lett.}
                 35 (1975) 338; {\em Phys. Rev.} D13 (1976) 3370.

\bibitem{FreTay} J.~Frenkel and J.C.~Taylor, {\em  Nucl. Phys.}
                 B116 (1976) 185.

\bibitem{APV}    D.~Amati, R.~Petronzio and G.~Veneziano,
                 {\em Nucl. Phys.} B146 (1978) 29.

\bibitem{MueCol} A.H. Mueller {\em Phys. Rev. } D20 (1979) 2037;\\
                 J.C.~Collins, {\em Phys. Rev. } D22 (1980) 1478;
                 in {\em Perturbative QCD}, ed. A.H. Mueller, 1989, p.~573.

\bibitem{Sen1} A.~Sen, {\em Phys. Rev. } D24 (1981) 3281.

\bibitem{Sen2} A.~Sen, {\em Phys. Rev. } D28 (1983) 860.


\bibitem{Bot}  G.~Sterman, {\em Nucl. Phys. } B281 (1987) 310; \\
               J.~Botts and G.~Sterman, {\em Nucl. Phys. } B325 (1989) 62.

\bibitem{Mag} L.~Magnea and G.~Sterman, {\em Phys. Rev. } D42 (1990) 4222.

\bibitem{SterLaen}
              H.~Contopanagos, E.~Laenen and G.~Sterman, 
              {\em Nucl. Phys. } B484 (1997) 303; \\
              N.~Kidonakis, G.~Odereda and G.~Sterman,
              {\em Nucl. Phys. } B531 (1998) 365; \\
              N.~Kidonakis, E.~Laenen, S.~Moch and R.~Vogt,
              Preprint TTP01-03, hep-ph/0105041.

\bibitem{BenSmi} M.~Beneke and V.A.~Smirnov, {\em Nucl. Phys.} B522 (1998) 321.

\bibitem{SmiRak} V.A.~Smirnov and E.R. Rakhmetov,
                 {\em Teor. Mat. Fiz.} 120 (1999) 64;\\
                 V.A.~Smirnov, {\em Phys. Lett. } B465 (1999) 226.

\bibitem{dimreg} G.~'t~Hooft and M.~Veltman, {\em Nucl.~Phys.}
                 B44 (1972) 189;\\
                 C.G.~Bollini and J.J.~Giambiagi,
                 {\em Nuovo Cim.} 12B (1972) 20.

\bibitem{Ste} G.~Sterman {\em Phys. Rev. } D17 (1978) 2773;\\
              S.~Libby and G.~Sterman {\em Phys. Rev. } D18 (1978) 3252;\\
              A.H.~Mueller, {\em Phys. Rep.} 73 (1981) 35.

\bibitem{MMN} T.~Matsuura, S.C.~van der Marck and W.L.~van Neerven,
              {\em Nucl. Phys.} {B319} (1989) 570;\\
              G.~Kramer and B.~Lampe,  {\em Z. Phys.} C34 (1987) 497.

\bibitem{KodTre} T.~Kodaira and L.~Trentedue,
                 {\em Phys. Lett.} B112 (1982) 66;\\
                 T.H.~Davies and W.J.~Stirling,
                 {\em Nucl. Phys.} {B244} (1984) 337;\\
                 G.P.~Korchemsky and A.V.~Radyushkin,
                 {\em Nucl. Phys.} {B283} (1989) 342.

\bibitem{Cat} S.~Catani, {\em Phys. Lett.} B427 (1998) 161.

\bibitem{AGOT} C.~Anastasiou, E.W.N.~Glover, C.~Oleari,
               M.E.~Tejeda-Yeomans,  {\em Nucl. Phys.} B601 (2001) 341.

\bibitem{BHM}  W.~Beenakker, W.~Hollik and S.C.~Van der Marck,
               {\em Nucl. Phys.} {B365} (1991) 24.

\end{thebibliography}
\end{document}